# Non-reciprocal charge transport in an intrinsic magnetic topological insulator MnBi$_2$Te$_4$

Zhaowei Zhang[1*], Naizhou Wang[1*], Ning Cao[2], Aifeng Wang[2], Xiaoyuan Zhou[2], Kenji Watanabe[3], Takashi Taniguchi[4], Binghai Yan[5†], Wei-bo Gao[1,6†]

[1] *Division of Physics and Applied Physics, School of Physical and Mathematical Sciences, Nanyang Technological University, Singapore 637371, Singapore*

[2]*Low Temperature Physics Laboratory, College of Physics and Center for Quantum Materials and Devices, Chongqing University, Chongqing 401331, China*

[3]*Research Center for Functional Materials, National Institute for Materials Science, Tsukuba, Japan*

[4]*International Center for Materials Nanoarchitectonics, National Institute for Materials Science, Tsukuba, Japan.*

[5]*Department of Condensed Matter Physics, Weizmann Institute of Science, Rehovot 7610001, Israel*

[6]*The Photonics Institute and Centre for Disruptive Photonic Technologies, Nanyang Technological University, Singapore 637371, Singapore*

*These authors contribute equally to this work.

†Corresponding author. Email: binghai.yan@weizmann.ac.il, wbgao@ntu.edu.sg

**Symmetries, quantum geometries and electronic correlations are among the most important ingredients of condensed matters, and lead to nontrivial phenomena in experiments, for example, non-reciprocal charge transport. Here we report the non-reciprocal charge transport in the intrinsic magnetic topological insulator MnBi$_2$Te$_4$. The current direction relevant resistance is observed at chiral edges, which is magnetically switchable, edge position sensitive and septuple layer number controllable. Applying gate voltage can effectively tune the non-reciprocal resistance. The observation and manipulation of non-reciprocal charge transport indicate the fundamental role of chirality in charge transport of MnBi$_2$Te$_4$, and pave ways to develop van der Waals spintronic devices by chirality engineering.**

Van der Waals materials provide an interesting platform to study the intertwined magnetism and band topology[1,2]. In particular, quantum anomalous Hall effect in van der Waals materials, such as intrinsic magnetic topological insulator MnBi$_2$Te$_4$(MBT)[3-6], twisted graphene[7] and twisted transition metal dichalcogenides[8] systems, shows higher critical temperature than that in magnetically doped topological insulators[9,10]. The chiral edge transport channels emerge at the Chern insulator states. Besides the quantum anomalous Hall effect, the chirality of the edge transport, as well as the magnetization, play an essential role in dissipative transport regimes, for instance, non-reciprocal charge transport behaviours[2,11-14].

Non-reciprocal resistance, manifesting as the resistance difference between positive and negative current, is the central process to convert an oscillating electromagnetic field to a direct current, in other words, rectification. The best known non-reciprocal resistance occurs in semiconducting p-n junctions. While working in microwave and terahertz frequency, however, the limitation of threshold voltage and operation frequency leads to the low efficiency of traditional semiconducting p-n junction-based rectifiers[15]. The demand of low-power, high-frequency rectifiers inspires studies on non-reciprocal charge transport in new material systems, such as non-centrosymmetric crystals[16-20], topological insulators[21-23], magnet/superconductor interfaces[24], topological insulator/superconductor interfaces[25] and magnet/topological insulator interfaces[11,26,27]. Especially, a large non-reciprocal charge transport mediated by quantum anomalous Hall edge states has been observed in magnetically doped topological insulator[22]. As compared to traditional magnetically doped topological insulators, the intrinsic magnetic topological insulator MnBi$_2$Te$_4$ doesn't introduce the impurities and removes the need of precise control of the ratio of different element species, and therefore promises high-quality devices and more robust topological properties.

In our current manuscript, we have demonstrated non-reciprocal charge transport in an intrinsic magnetic topological insulator MnBi$_2$Te$_4$. MnBi$_2$Te$_4$ is an A-type antiferromagnet, where each Mn$^{2+}$ ion contributes the magnetic moment of $\sim 5\mu_B$. The magnetic moment of the Mn$^{2+}$ is ferromagnetically aligned within the septuple layer (SL), but antiferromagnetically aligned between two adjacent septuple layers. The nature of van der Waals structure provides the possibility of multi-dimensional control toolbox including gate voltage, magnetic field and septuple layer numbers. The observation of tunable non-reciprocal resistance may help to develop multifunctional spintronic devices, such as magnetic switchable diodes and high-frequency rectification.

First, we show the characterization of MnBi$_2$Te$_4$ devices. The device structure is shown in Figure 1a. In details, MnBi$_2$Te$_4$ thin flakes are obtained by Al$_2$O$_3$ assisted exfoliation[3,28]. The sample thickness is determined by optical contrast as shown in Figure S1. We have chosen two 5-SL (Device 1 and Device 2) and one 4-SL (Device 3) MnBi$_2$Te$_4$ in our measurements. In the main text of this Letter, we show results of Device 1 and Device 3. Au electrodes are thermally deposited through stencil masks. This method avoids the problems of the chemical or ambient exposure and ensures the high device quality. The Si/SiO$_2$ bottom gate or the graphite/hBN top gate works as gate. We measure the longitudinal resistance of a 5-SL MnBi$_2$Te$_4$ as a function of temperature as shown in Figure 1b. The Néel temperature is around 23 K, which is indicated by the resistance peak. The temperature dependent resistance of other two devices are shown in Figure S2a and d. Three devices display similar Néel temperature.

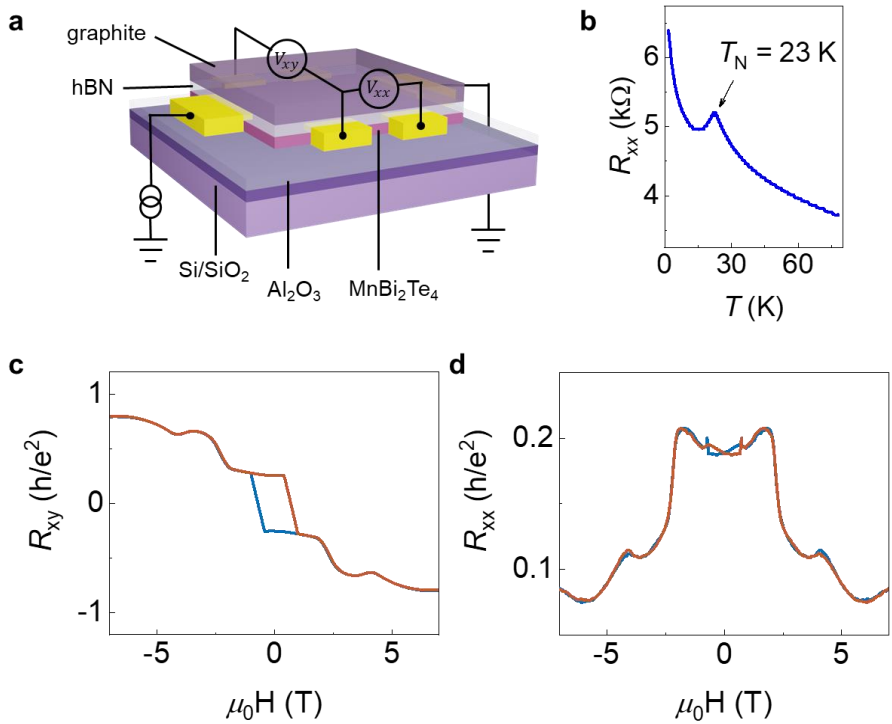

**Figure 1: The characterization of MnBi$_2$Te$_4$ devices. a**, The schematic structure of the MnBi$_2$Te$_4$ device. The MnBi$_2$Te$_4$ thin flakes are obtained by Al$_2$O$_3$ assisted exfoliation method. Au electrodes are thermally deposited through stencil masks. Si/SiO$_2$ and graphite/hBN work as bottom and top gate, respectively. **b**, The temperature dependent resistance of a 5-SL MnBi$_2$Te$_4$ (Device 1) device. The Néel temperature is indicated by a resistance peak around 23 K. **c** and **d,** The transverse and longitudinal resistance as a function of the magnetic field at 1.7 K for Device 1. A top gate voltage of -2 V is applied to tune the Fermi energy to the charge neutrality point.

To reveal the chiral edge states in MnBi$_2$Te$_4$, we tune the Fermi level to the charge neutrality point (CNP). 5-SL MnBi$_2$Te$_4$ shows an uncompensated magnetism under zero external magnetic field as the temperature is below the Néel temperature, as shown in Figure 1c, 1d. We also show the fully compensated antiferromagnetic states of a 4-SL MnBi$_2$Te$_4$ device in Figure S4d. We note that in the ideal quantum anomalous Hall state, only chiral states contribute to the charge transport and the longitudinal resistance becomes zero. However, longitudinal resistance in our MnBi$_2$Te$_4$ samples does not vanish. For our Device 2 and Device 3, the nearly quantized Hall resistance of 0.977 h/e$^2$ and 0.945 h/e$^2$ have been achieved (Figure S2).

Next, we study the non-reciprocal charge transport in MnBi$_2$Te$_4$ devices. As shown in Figure 2a, when the current is injected in opposite directions, charge carriers at the same edge are scattered by the chiral state in different ways. This leads to a difference in resistance for positive and negative current, so-called nonreciprocal charge transport behaviours. Here, we adopt the phenomenological model[29] to describe the current relevant resistance:

$$V_{xx} = iR_{xx} = iR_0 + \gamma R_0 i^2 (\widehat{M} \times \widehat{P}) \cdot \hat{\imath}, \qquad (1)$$

where $R_0$ is the resistance that does not change with the current, $\gamma$ is the constant that characterizes the strength of the non-reciprocal charge transport effect, $\widehat{M}$ is the magnetization direction of the MnBi$_2$Te$_4$, $\widehat{P}$ is the edge charge dipole which is opposite on two edges, and $\hat{\imath}$ is the current direction. We study the non-reciprocal charge transport by measuring the resistance difference between positive and negative currents in a 5-SL MnBi$_2$Te$_4$ as described in Figure 2a. The 10 µA and -10 µA DC current is injected to the device respectively, and we measure the longitudinal resistance at the right edge at 11 K. A top gate voltage of -2 V is applied, at which the Fermi level is around the CNP. We note that the chirality of the edge transport channels can be controlled by magnetic moments. In Figure 2b, we plot the resistance difference as a function of the magnetic field. The resistance difference $\Delta R$ is defined as $R_L(10\ \mu A) - R_L(-10\ \mu A)$ and it shows the difference in resistance between positive and negative current. The value is antisymmetrized for ***M***. We also note a finite magnitude without external magnetic field, which is due to the uncompensated magnetic moments.

Since the non-reciprocal response are originated from the quadratic term of current, we adopt the second-harmonic voltage measurements with small AC driven current $I^{RMS} = 5\ \mu A$ to obtain better signal-to-noise ratio and rule out the heating effect and other high order effects. With the base frequency of the driven current set as 17.777 Hz, the voltage drop is written as:

$$V_{xx} = \sqrt{2}\sin(\omega t)R_0 + \gamma R_0 i^2 (1 - \cos(2\omega t))(\widehat{M} \times \widehat{P}) \cdot \hat{\imath}. \qquad (2)$$

The voltage with the frequency of 35.554 Hz reveals the strength of the non-reciprocal charge transport effect. The non-reciprocal resistance as a function of magnetic field is shown in Figure 2c, which shows the same behaviour as that of DC measurements. The $R_R^{2\omega} - H$ loop exhibits several kinks that indicate the phase transition between different magnetic states, which agrees with the 1st order Hall measurements. We show the non-reciprocal resistance in Device 2 in Figure S5.

We then focus on the effect of edge-state chirality on the non-reciprocal charge transport. The magnetic field dependent non-reciprocal resistance at the left edge is presented in Figure 2d, which shows opposite trend compared to that of the right edge and suggests that the inversion symmetry of edge transport is broken. Apart from fully spin polarized states, nonreciprocal charge transport behaviours are also observed without external magnetic field. Owing to the opposite net magnetization, in other words, opposite chirality of the edge transport channels, at the spin ↓↑↓↑↓ state and spin ↑↓↑↓↑ state, the $R_L^{2\omega} - H$ curves show a hysteresis.

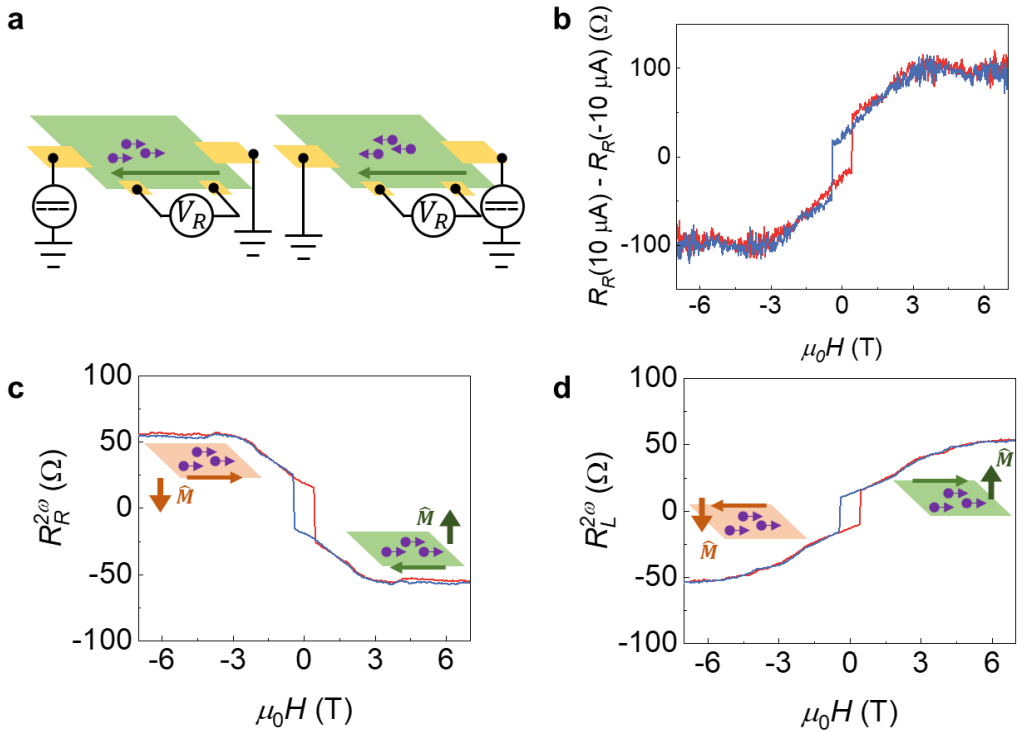

**Figure 2: Non-reciprocal charge transport in a 5-SL MnBi$_2$Te$_4$. a**, Schematic illustrations of current direction dependent backscattering of edge transport. **b**, Resistance difference between positive and negative current as a function of the magnetic field. The measurement is performed at 11 K with top gate voltage of -2 V. **c** and **d**, The non-reciprocal resistance measured at the right edge and left edge, respectively. The measurements are performed at 11 K with top gate voltage of -2 V. Insets show schematic illustrations of the interplay between chiral edge states and 2D bulk transport channels (e.g., surface states) at different magnetic states.

Magnetic moments of 4-SL $MnBi_2Te_4$ are fully compensated under zero magnetic field, for which the non-reciprocal charge transport in 4-SL $MnBi_2Te_4$ shows different behaviours compared with that in 5-SL $MnBi_2Te_4$. In Figure 3a and b, we show the field dependent non-reciprocal resistance of 5-SL and 4-SL $MnBi_2Te_4$ devices, respectively. The non-reciprocal resistance for both devices are measured at the right edge with slightly electron doping. We find in high-magnetic-field regime, non-reciprocity of 5-SL $MnBi_2Te_4$ and 4-SL exhibit the same order of magnitude. However, under zero magnetic field, non-reciprocal resistance nearly vanish in 4-SL $MnBi_2Te_4$ devices. This is consistent with the phenomenological model for $M = 0$ and the current relevant term vanishes. The vanished non-reciprocity with zero magnetization can be further confirmed by measuring the non-reciprocal resistance at paramagnetic states. Increasing the temperature above the Néel temperature results in the breakdown of chiral edge states even under high magnetic field, leading to vanished non-reciprocal resistance, which is shown in Figure S7.

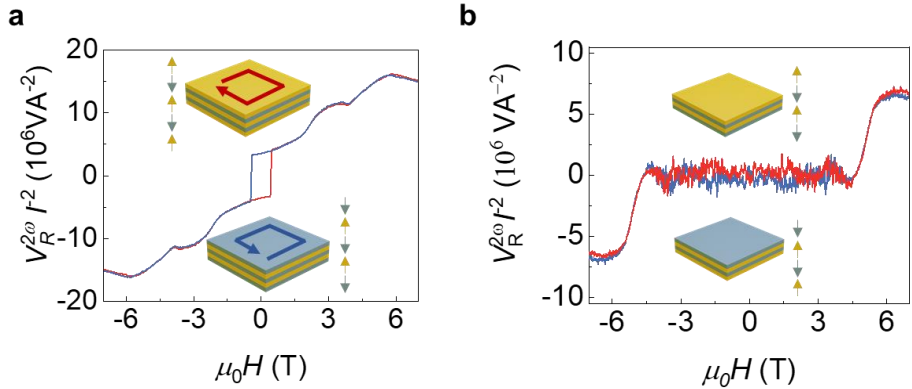

**Figure 3: Non-reciprocal charge transport in 5-SL $MnBi_2Te_4$ and 4-SL $MnBi_2Te_4$. a**, Magnetic field dependent non-reciprocal resistance for a 5-SL $MnBi_2Te_4$ device. A top gate voltage of 3 V is applied, and the sample is slightly electron doped. Insets show the schematic illustrations of two magnetic states under zero magnetic field. **b**, Magnetic field dependent non-reciprocal resistance for a 4-SL $MnBi_2Te_4$ device. The AC driven current $I^{RMS} = 1 \mu A$ is adopted in measurements. The measurements are performed at the temperature of 11 K. A top gate voltage of 3 V is applied, and the sample is slightly electron doped. Insets show schematic illustrations of two magnetic states under zero magnetic field.

Finally, we demonstrate the gate tunability of the non-reciprocal charge transport in $MnBi_2Te_4$. In Figure 4, we show the gate dependent $R_L^{2\omega}$ and $R_R^{2\omega}$ at 11 K with various top gate voltage. We find that both the magnitude and sign of the $R_{L/R}^{2\omega}$ can be tuned by applying gate voltage. In high field regime, $R_L^{2\omega}$ at the spin state ↑↑↑↑↑ is tuned from positive to negative when gate voltage is scanned from -5V to 5V, while $R_R^{2\omega}$ shows the opposite trend. The gate tunability is also valid under zero magnetic field, where the loop chirality is switched while scanning top gate voltage from -5V to 5V. We note the gate voltage does not switch the ↑↓↑↓↑ and ↓↑↓↑↓ states as we discussed in Supplementary Note 5. Therefore, it is the change of Fermi energy that affects the non-reciprocal charge transport.

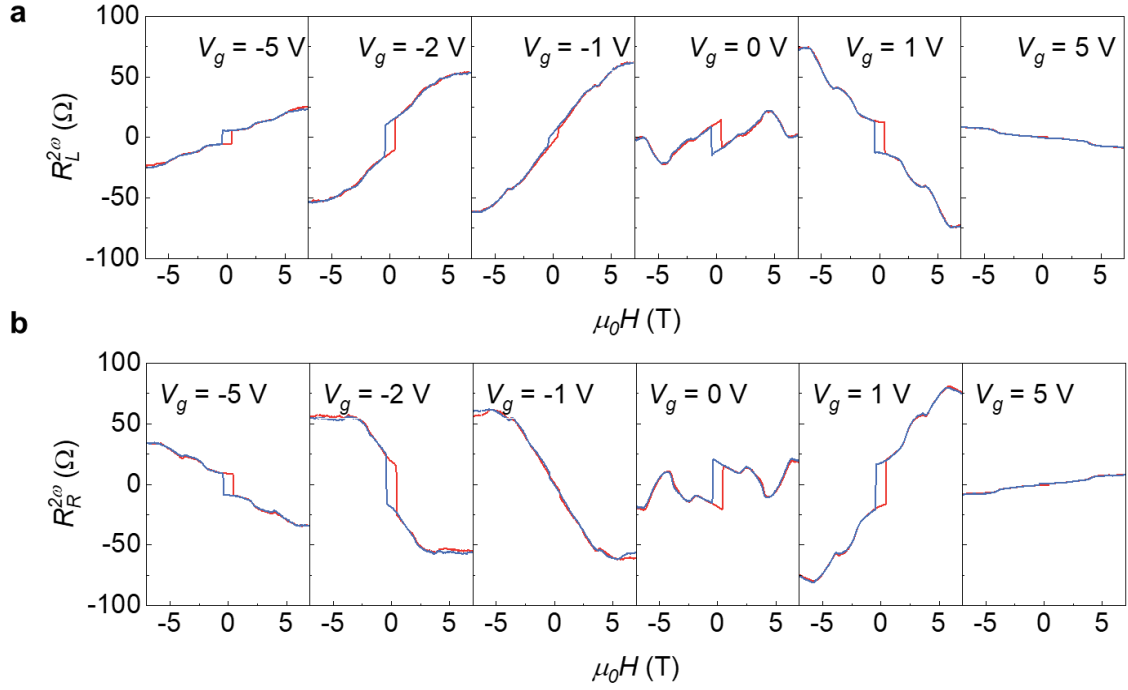

**Figure 4: Non-reciprocal resistance at various top gate voltages. a**, Non-reciprocal resistance as a function of magnetic field measured at the left edge. **b**, Non-reciprocal resistance as a function of magnetic field measured at the right edge.

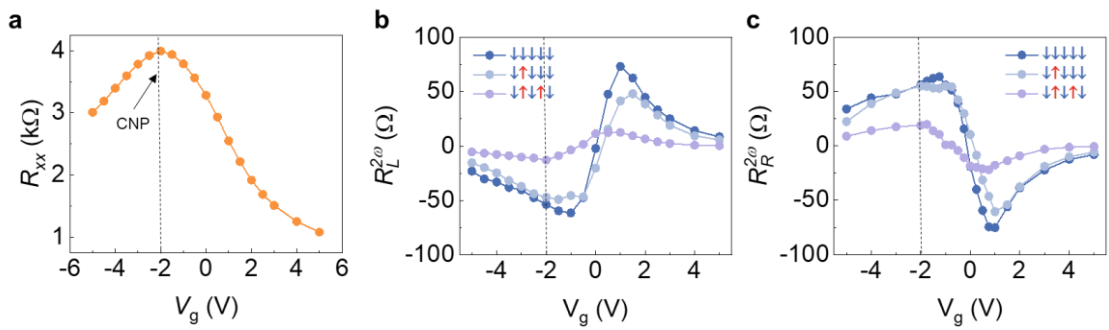

**Figure 5: Gate tunability of non-reciprocal resistance at different magnetic states in 5-SL MnBi$_2$Te$_4$. a**, Longitudinal resistance of the 5-SL MnBi$_2$Te$_4$ as a function of the top gate voltage under zero magnetic field. The measurement is carried out at 11 K. The resistance shows a maximum at the charge neutrality point with the top gate voltage of -2 V. **b** and **c**, Non-reciprocal resistance measured at left edge and right edge of the 5-SL MnBi$_2$Te$_4$, respectively. Different curves show the non-reciprocal resistance at spin states of ↓↓↓↓↓, ↓↑↓↓↓ and ↓↑↓↑↓.

To provide more insights into the effect of gate voltage on the nonreciprocal charge transport, we summarize the gate dependence of longitudinal resistance and non-reciprocal resistance in Figure 5. As shown in Figure 5a, the longitudinal resistance under zero magnetic field shows the maximum at the top gate voltage of -2 V. We also show sign reversal of the $R_L^{2\omega}$ and $R_R^{2\omega}$ while scanning gate voltage at different magnetic states in Figure 4b and c.

The remarkable features of non-reciprocal resistance can be rationalized by a simple band structure scenario on the QAHE edge. At the CNP, we should note that our devices deviate slightly from the fully insulating state and $R_{xy}$ being smaller than $h/e^2$ (see Figure 1c). Therefore, one chiral edge state and trivial edge states (including subbands from the 2D bulk) coexist, from which the local edge transport is detected by two voltage electrodes. Apart from MnBi$_2$Te$_4$, prior studies on magnetically doped topological insulators show the small finite longitudinal resistance in presence of nonchiral edge states and residual bulk states[30]. In the ideal QAHE case, the pure chiral edge state would not lead to nonreciprocal transport, since its $R_{xx} = 0$. In the weakly doped QAHE which is the case in present work, chiral edge state hybridizes strongly with trivial edge states, leading to asymmetric dispersion between opposite momenta, i.e., different magnitudes of Fermi velocities along opposite directions. This velocity asymmetry coincides with the fact that both inversion symmetry and time-reversal symmetry are broken on the edge. Suppose a finite relaxation time, the direction-dependent mean free path comes with direction-dependent Fermi velocity and eventually leads to direction-dependent resistance. Ref. 15 reveals that the induced resistance change is maximized near the band edge of trivial states, explaining the large nonreciprocal effect near the CNP. Further, the velocity asymmetry flips order between conduction and valence bands, or when reversing the magnetism or changing edge sides, rationalizing related experimental sign changes.

In summary, we have demonstrated the non-reciprocal charge transport in an intrinsic topological insulator MnBi$_2$Te$_4$. The septuple layer dependent non-reciprocity is magnetically controllable and edge position sensitive. Meanwhile, the observed non-reciprocal resistance can be tuned by gate voltage. We ascribe the observed non-reciprocal resistance to the interaction between chiral edge states and dissipative edge states. Our finding paves the way to build next-generation spintronic devices though chirality engineering.

**Methods:**

**Device Fabrication.** High-quality MnBi$_2$Te$_4$ crystals are grown by flux methods. MnBi$_2$Te$_4$ thin flakes are obtained by Al$_2$O$_3$ assisted exfoliation method. We determine the thickness of MnBi$_2$Te$_4$ thin flakes by optical contrast of captured optical images of MnBi$_2$Te$_4$/Al$_2$O$_3$ on PDMS stamps. MnBi$_2$Te$_4$ thin flakes, as well as the Al$_2$O$_3$ film are then transferred on Si/SiO$_2$ substrates. Au electrodes are thermally deposited through stencil masks. Finally, the hBN and graphite are transferred on top of the selected MnBi$_2$Te$_4$ flake. The fabrication processes before covering hBN and graphite are performed in a N$_2$ filled glove box.

**Electrical measurements.** DC transport measurements are performed by a current source (Keithley 6221) and a nanovoltmeter (Keithley 2182A). The top gate voltage and bottom gate voltage are applied by a dual-channel sourcemeter (Keithley 2636B). AC transport measurements are performed by a current source (Keithley 6221) and a lock-in amplifier (Zurich MFLI). All measurements are carried out with a Cryomagnetic cryostat.

## Supplementary Information

### Supplementary Note 1: Layer thickness characterisation

We have fabricated three MnBi$_2$Te$_4$(MBT) devices, including two 5-SL (Device 1 and Device 2) and one 4-SL (Device 3) MBT devices. The MBT thin flakes are obtained by Al$_2$O$_3$ assisted exfoliation method and the thickness of the thin flakes are determined by optical contrast[3,28]. The transmission is defined as $I_{sample}/I_{substrate}$, where $I_{sample}$ and $I_{substrate}$ are the intensity of the transmission through the sample and substrate respectively, in the green channel of the optical image.

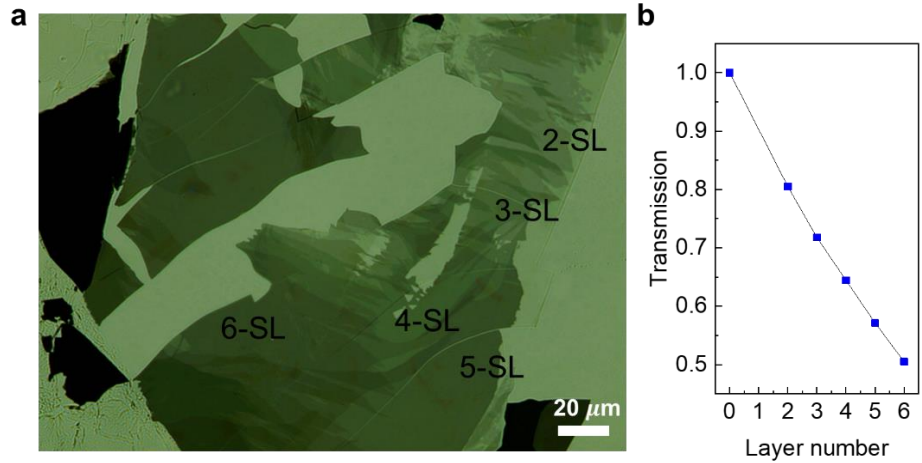

**Figure S1: Exfoliation of MBT with Al$_2$O$_3$ assisted exfoliation method. a**, Optical image of the MBT thin flakes on polydimethylsiloxane(PDMS)/Al$_2$O$_3$ substrates. The layer number is labelled. **b**, Transmission as a function of the number is SLs.

### Supplementary Note 2: Quantum anomalous Hall(QAH) states in Device 2 and Device 3

Figure S2 shows the quantum anomalous Hall states of Device 2 and Device 3. Both 5-SL MBT and 4-SL MBT shows the Néel temperature of 23 K, which is determined by the resistance peak in Figure S2a and d.

We also investigate the effect of gate voltage on the Hall resistance and longitudinal resistance under -7 T magnetic field at 1.7 K. Figure S2a and b show the gate dependence of Device 1 and Device 3. At the charge neutrality point, the Hall resistance reaches the maximum and the longitudinal resistance reaches the minimum.

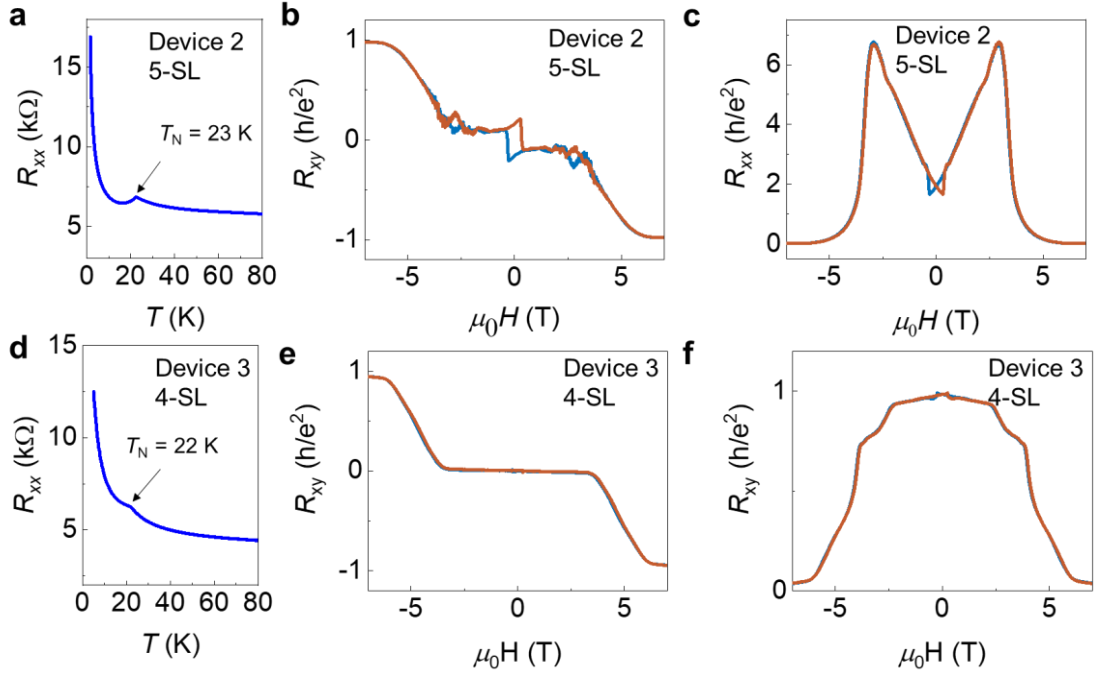

**Figure S2: The quantum anomalous Hall states in 5-SL and 4-SL MBT devices. a**, **b** and **c** show the temperature dependent resistance, field dependent Hall resistance and longitudinal resistance of Device 2 (5-SL). **d**, **e,** and **f** show the temperature dependent resistance, field dependent Hall resistance and longitudinal resistance of Device 3 (4-SL)

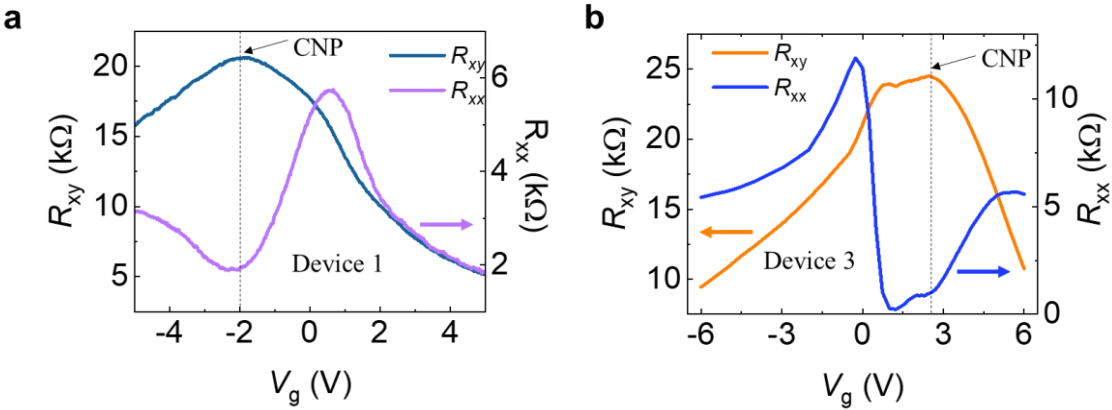

**Figure S3: The gate dependence of 5-SL and 4-SL MBT devices. a** The gate voltage dependent Hall and longitudinal resistance of Device 1 under -7 T magnetic field at 1.7 K. **b** The gate voltage dependent Hall and longitudinal resistance of Device 3 under -7 T magnetic field at 1.7 K.

The temperature dependence of Device 1 and Device 3 are shown in Figure S4. Under -7 T out-of-plane magnetic field, increasing the temperature leads to the reduced Hall resistance and the longitudinal resistance is thermally activated following the relation, $R_{xx} \propto e^{\frac{-\Delta}{2k_B}}$, where $\Delta$ is the activation energy gap and $k_B$ is the Boltzmann constant. By linearly fitting the Arrhenius plot of $\ln(R_{xx})$ as a function of $\frac{1}{T}$, we estimated the $\Delta$ under -7 T magnetic field to be 0.944 meV in 5-SL and 0.779 meV in 4-SL MBT.

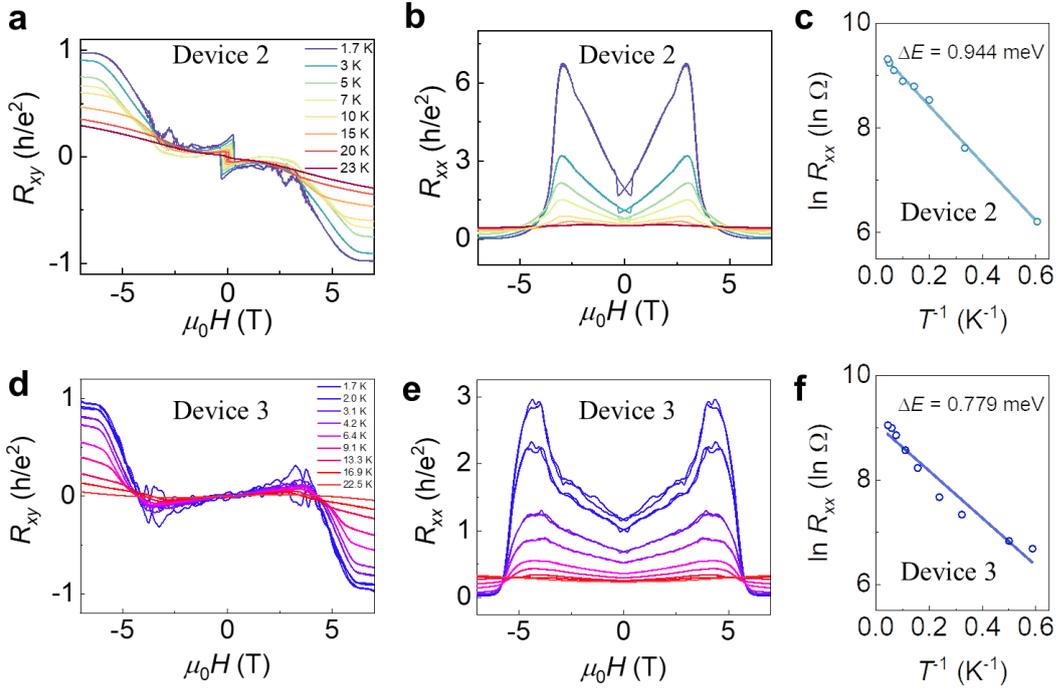

**Figure S4: Temperature dependence of Hall resistance and longitudinal resistance as a function of the magnetic field. a**, **b** and **c** The temperature dependent Hall resistance and longitudinal resistance of the 5-SL MBT (Device 2) and the estimated activation energy gap of 0.944 meV. The measurement is performed with the bottom gate voltage around the CNP of $V_g^{CNP}$=22 V. **d**, **e** and **f** The temperature dependent Hall resistance and longitudinal resistance of the 4-SL MBT (Device 3) and the estimated activation energy gap of 0.779 meV. The measurement is performed with the bottom gate voltage around the CNP of $V_g^{CNP}$=13 V.

## Supplementary Note 3: Non-reciprocal charge transport in Device 2

We have shown the gate dependent non-reciprocal charge transport in a 5-SL MBT (Device 1) in Figure 4 of the main text. Here we show the gate dependent non-reciprocal charge transport in another 5-SL MBT (Device 2). The AC current $I^{RMS}$ injected into the sample is 1 μA.

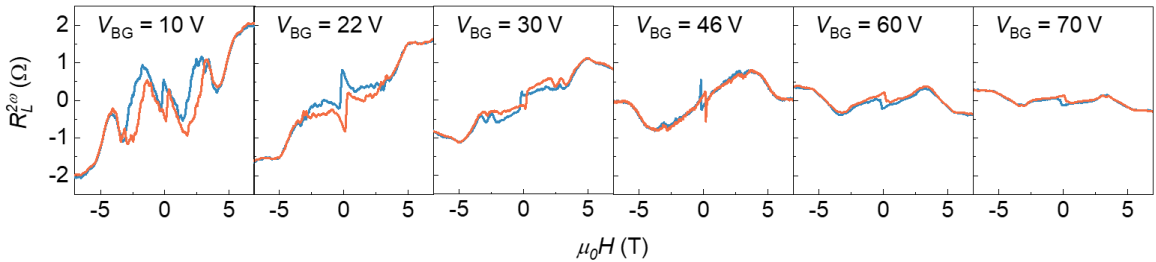

**Figure S5: Non-reciprocal charge transport of Device 2 (5-SL) at different back gate voltage.** Two electrodes at the left sides of the sample are used to measure the non-reciprocal resistance. The measurement is performed at 10 K with AC current $I^{RMS} = 1$ μA.

We also summarize the gate dependent non-reciprocal resistance under -7 T out-of-plane magnetic field in Figure S6. Under -7 T out-of-plane magnetic field, at the CNP with bottom gate voltage around $V_{BG}^{CNP} = 22$ V, the Hall resistance reaches the maximum and the longitudinal resistance reaches the minimum as shown in Figure S6 a. In Figure S6 b, we show the non-reciprocal resistance measured at the left edge of the Device 2.

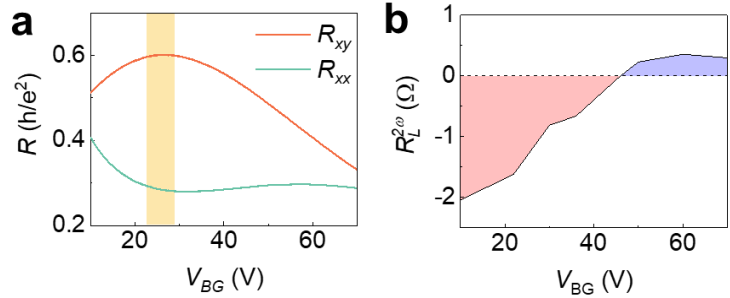

**Figure S6: The gate dependent non-reciprocal transport in 5-SL MBT (Device 2) a** Hall resistance and longitudinal resistance as a function of the bottom gate voltage. The measurement is performed at 10 K under -7 T magnetic field. The Hall resistance reached the maximum and the longitudinal resistance reaches the minimum at the CNP. **b** The non-reciprocal resistance at the left edge as a function of the bottom gate voltage.

## Supplementary Note 4: Vanished nonreciprocal charge transport at high temperature

We show the effect of temperature on the nonreciprocal resistance in Figure S6. In Device 3, the nonreciprocal resistance vanishes when the temperature is 30 K, which is higher than its Néel temperature. The vanished non-reciprocal resistance corresponds to magnetization $M \approx 0$. Under this condition, the current relevant term in Equation 1 of the main text vanishes.

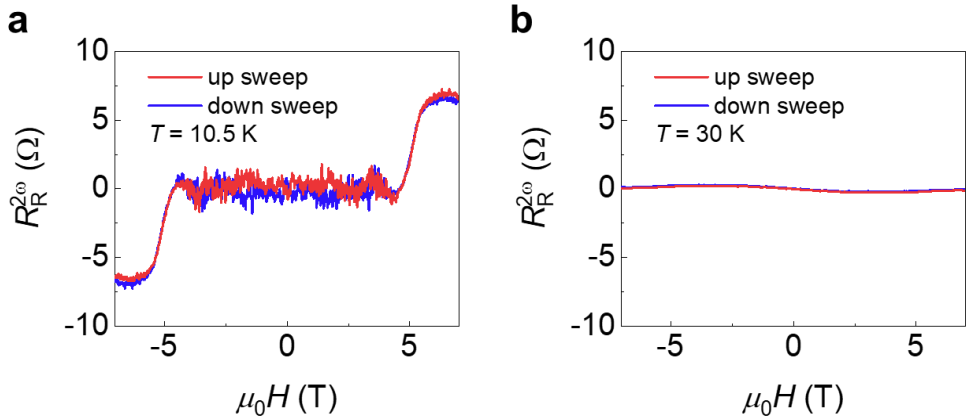

**Figure S7: The non-reciprocal resistance of 4-SL MBT device at different temperature a** and **b** The non-reciprocal resistance in Device 3 (4-SL) measured at 10.5K and 30 K, respectively. The measurements are performed with top gate voltage of 3 V.

## Supplementary Note 5: Magnetic states of 5-SL at different gate voltage

We study the effect of gate voltage on magnetic states in 5-SL MBT. Figure S7 shows field dependent Hall resistance at top gate voltage of -5 V, 0 V and 4 V of the 5-SL MBT (Device 1). We note the magnitude of the Hall resistance changes at different gate voltage, but the sign does not change.

This indicates the magnetic states of 5-SL MBT under zero magnetic field are independent to the gate voltage in our device.

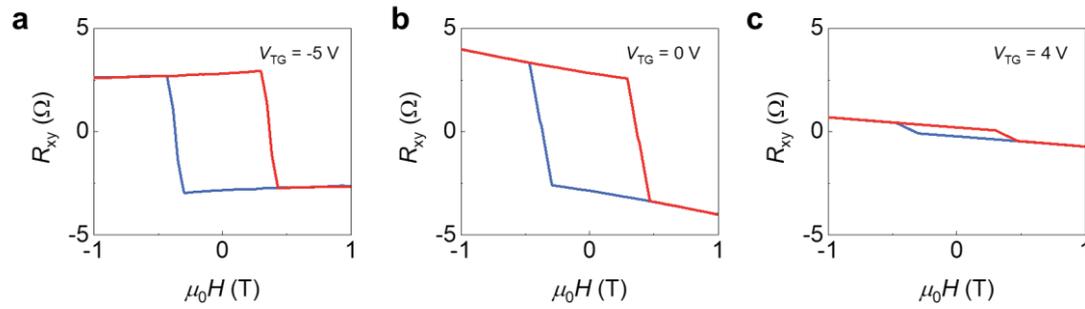

**Figure S8: The Hall resistance of 5-SL MBT device(Device 1) at different gate voltage**
Field dependent Hall resistance at top gate voltage of -5 V, 0 V and 4 V of the 5-SL MBT (Device 1).